\documentclass[10pt]{article}
\usepackage[small,bf]{caption}
\newsavebox{\PSLASH}
\sbox{\PSLASH}{$p$\hspace{-1.8mm}/}

\begin{document}
\title{On the Probability of Occurrence of Clusters in Abelian Sandpile Model }
\author{M. Moradi \footnote{e-mail: $mm_{-}moradi@mehr.sharif.edu$},
S. Rouhani \footnote{e-mail: $rouhani@ipm.ir$}
\\
Department of Physics, Sharif University of Technology,\\ Tehran,
P.O.Box: 11365-9161, Iran}
\date{}
\maketitle

\begin{abstract}
We have performed extensive simulations on the Abelian Sandpile
Model (ASM) on square lattice. We have estimated the probability
of observation of many clusters. Some are in good agreement with
previous analytical results, while some show discrepancies between
simulation and analytical results.
\end{abstract}

\section{Introduction}
\indent

 Abelian Sandpile Model (ASM) is a generalization of BTW
model, proposed by Bak, Tang and Wisenfeld \cite{b.BTW} for
understanding the mechanism of Self-Organized Criticality (SOC).
Although the physical relevance of ASM can be questioned, it is
the simplest and most studied one among all other SOC systems. It
is believed that, the model is an exactly solvable with all the
features of the SOC systems \cite{b.Dhar}. There are very good
reviews on this model, for example see \cite{b.Ivash,b.DharRev}.

BTW sandpile model is defined as a simple cellular automata on the
square lattice of size $L\times M$. On each site $i$ a variable
$h_{i}$ is defined as its height, taking values  \{$1, 2, 3, 4$\}.
System evolves in discrete time: at each step one site is picked
randomly and its height is increased by unity. If its height
becomes larger than $h_{c}=4$, this site is said to be unstable
and it relaxes by toppling. The process of toppling is defined as
follows: four grains leave the unstable site and each of its four
neighbors gets one grain. So the number of grains is conserved
during the toppling, except at the boundaries where some of the
grains leave the system. If there is any unstable site remaining,
it will also topple.

The process of toppling could be stated in another way. If the
site $i$ becomes unstable, $h_j$ will be decreased by amount of
$\Delta_{ij}$, that is $h_{j}\rightarrow h_{j} - \Delta_{ij}$
where
\begin{eqnarray}\
    \Delta_{ij}=\left\{%
\begin{array}{ll}
    4, & \hbox{ $i=j$;} \\
    -1, & \hbox{ $i, j$ are neighbors;} \\
    0, & \hbox{ otherwise.} \\
\end{array}%
\right.
\end{eqnarray}

After a while the system reaches a steady state, in which it shows
SOC. It has been shown that in this state, the number of different
configurations the system accepts is $\det \Delta$ and the
probability of all of them are the same. These configurations are
named recurrent configurations in contrast with the transient ones
which can only appear in the first steps of evolution, where the
system has not yet reached the steady state.

\begin{figure}[h]
\begin{picture}(150,150)(0,0)
\includegraphics{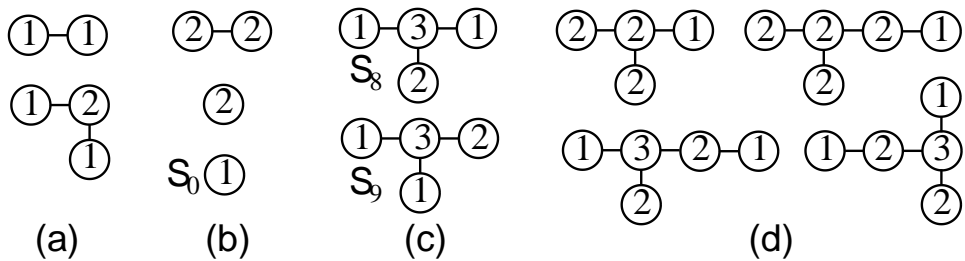}
\end{picture}
\caption{ Some different types of clusters: (a) Forbidden
Sub-Configurations. (b) Allowed Clusters ($S_{0}$ is also Weakly
Allowed Cluster). (c) $S_{8}$, $S_{9}$ have exactly the same
analytical values but different simulation results. (d) Some of
weakly allowed clusters which are in disagreement with analytical
results.} \label{f.Clusters}
\end{figure}

The question is, what is the probability of finding one given
cluster in SOC mode? It is possible to show that some
sub-configurations are disallowed in SOC mode (e.g. clusters in
fig.~\ref{f.Clusters}-a). There is an easy way to test whether a
given cluster is forbidden or allowed, so the probability of some
clusters are exactly equal to zero. Dhar and Majumdar calculated
the probability of finding some allowed clusters using a beautiful
idea \cite{b.DharMajum}. This method is appropriate only for
weakly allowed clusters. These clusters have the property that
decreasing the height of any of their sites, makes them forbidden
(such as $S_{0}$ in fig.~\ref{f.Clusters}-b and all clusters in
fig.~\ref{f.Clusters}-c and fig.~\ref{f.Clusters}-d). The first
cluster calculated analytically in \cite{b.DharMajum} was $S_0$:
\begin{equation}\label{e.P0}
P(S_0)=\frac{2}{\pi^{2}}-\frac{4}{\pi^{3}}=0.0736362...
\end{equation}
A table of some other analytical results can be found in
\cite{b.Ruelle}. It is obvious that $S_{10}$ and $S_{11}$ in ref.
\cite{b.Ruelle} are not weakly allowed clusters and one can not
use this method for them.

Yet there are many other clusters that are not included in this
set of clusters, so their probabilities are still undetermined. It
seems that some generalization is required for exact calculation
of probability of these clusters. Some attempts have been made in
this direction, for example in \cite{b.Priz} all height
probabilities have been calculated.

Several works have been done based on analytical approach in
\cite{b.DharMajum} such as calculation of correlation functions
\cite{b.DharMajum}, probability of clusters and correlation
functions in boundaries \cite{b.Ivash1,b.Bran} and scaling fields
\cite{b.Ruelle}. Nevertheless, we focus only on the first problem:
"the probability of weakly allowed clusters" because it contains
all foundations of analytical approach throughout the literature.

\section{Simulation results}
\indent

 The first simulations were done on small lattices and for
simple clusters such as $S_0$. Extensive simulations for the
lattice of size $672\times672$ were undertaken by Manna in
\cite{b.Manna}, who found $P(S_{0})=0.0736\pm0.003$. Grassberger
and Manna also performed simulations on some larger lattices, but
not with sufficiently high statistics \cite{b.ManGras}.

Although these results are in agreement with analytical result in
eq. (\ref{e.P0}), it seems  more clusters should be tested with
various sizes and forms. It is also needed to perform more precise
simulation using larger lattices and more samples.

\begin{figure}[h]
\begin{picture}(155,155)(0,0)
\includegraphics{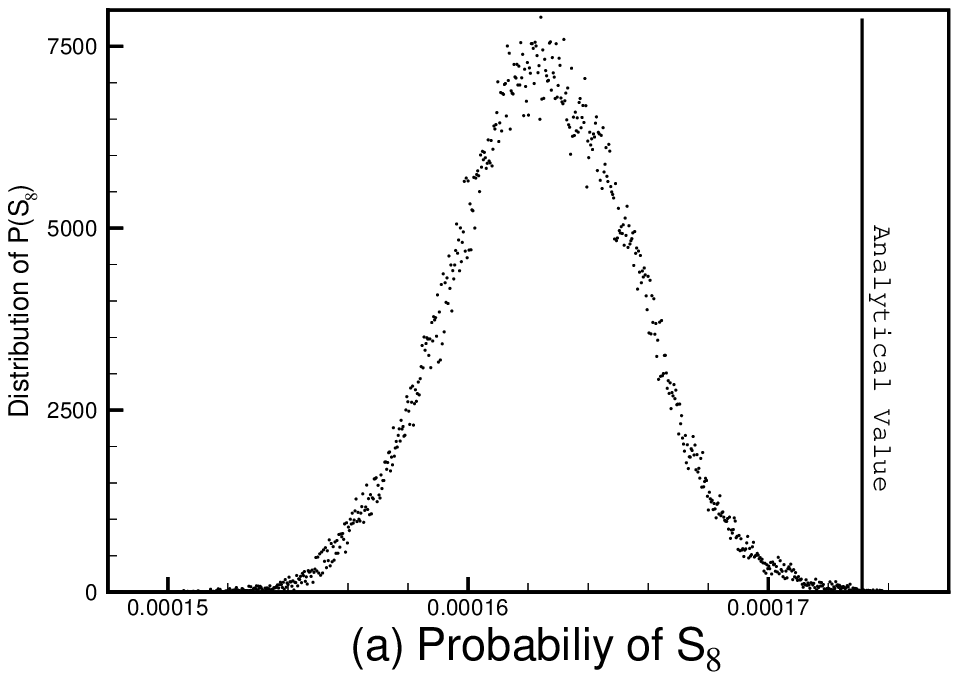}
\includegraphics{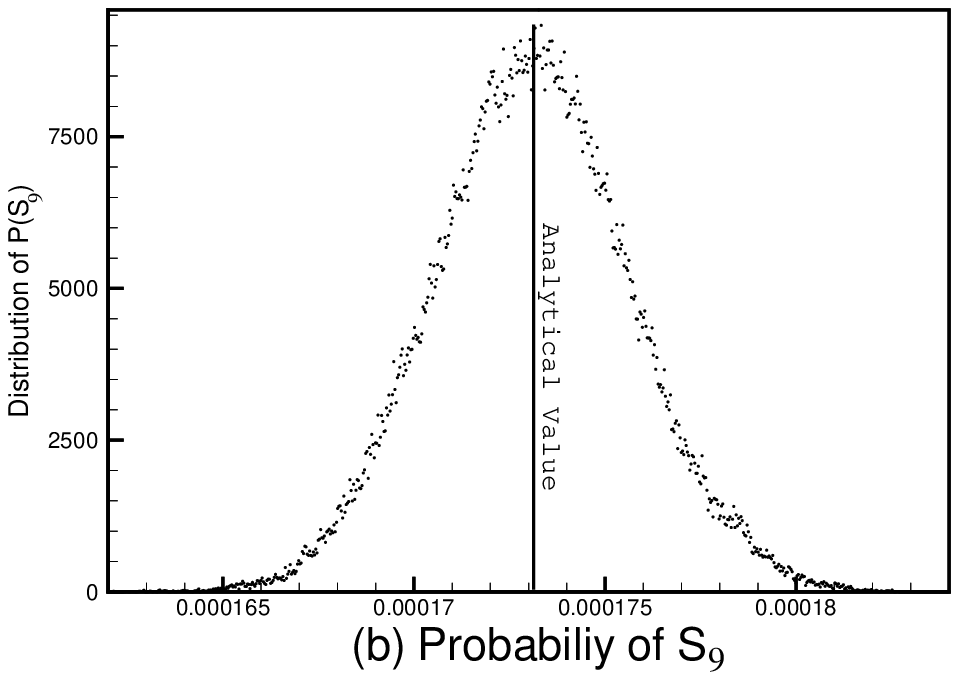}
\end{picture}
\caption{Distribution of probability of clusters over 2,000,000
samples on lattice of size $2000\times2000$: (a) $P(S_{8})$ is in
disagreement with analytical value (thick bar). (b) $P(S_{9})$ is
in good agreement with analytical value.}\label{f.Curves}
\end{figure}

We first started with a lattice of size $500\times500$ and
extended it to a $2000\times 2000$ one and  measured the
probability of clusters introduced in the table of paper
\cite{b.Ruelle} for 2,000,000 samples. sampling is started when
the system reaches SOC using the burning test \cite{b.Burn}.  Each
new sample is obtained by adding a grain of sand and allowing the
system to relax. This process is repeated for 2,000,000 times and
we find a distribution function for probability of appearance of
any given cluster. The result is typically a Guassian distribution
(see fig.~\ref{f.Curves}).

Statistical analysis shows that for one of the clusters ($S_{8}$)
there is a deviation from analytical result(fig.~
\ref{f.Curves}-a)\footnote{In lattice of size $500\times500$ , the
deviation was not so clear, though one could guess that there is
something wrong.}. To confirm that the deviation does exist and is
not a simulation error, we have done the same simulation for the
cluster $S_{9}$, which topographically is the same as $S_{8}$.
Also probability of $S_{9}$ is analytically the same as that of
$S_{8}$ and simulation confirms its analytical result without any
remarkable deviation as fig.~\ref{f.Curves}-b.

It is also noticeable that the quantity studied by analytical
approach, is an average over all recurrent configurations.
Standard deviations in fig.~\ref{f.Curves}-a and
fig.~\ref{f.Curves}-b are the error in each individual measurement
of the sample, so the more appropriate quantity to stand for error
bar is the standard error of the mean ($\sigma_m$). In our
simulations it was generally much less than the error of precision
concerning each individual sample which is equal to $\frac{1}{N}$
where $N$ is the number of all attempts to find the specific
cluster in each sample. This quantity, therefore, is the best one
for estimating the error.

The simulation results for the clusters $S_{8}$ and $S_{9}$ were
obtained :
\begin{eqnarray}
P_{2000\times2000}(S_{8}) = 0.000162467 \pm 0.000000031 & \sigma = 0.000003072 & \sigma_{m} = 0.0000000022 \\
P_{2000\times2000}(S_{9}) = 0.000173098 \pm 0.000000031 & \sigma =
0.000002558 & \sigma_{m} = 0.0000000019
\end{eqnarray}
to be compared with the analytical result :
\begin{equation}
 P_{\rm Analytic}(S_{8}) = P_{\rm Analytic}(S_{9}) = 0.000173106... \\
\end{equation}

We repeated the simulation on large lattices of size
$500\times500$, $1000\times1000$ and $2000\times2000$ for all
clusters of $size\leq6$. We found that there are more clusters in
disagreement with the analytical results (e.g. clusters in fig.~
\ref{f.Clusters}-d), though they are in a minority.

Repetition of simulation confirms our results and shows that there
is a noticeable disagreement between analytical result and
simulation. The origin of this disagreement remains to be
understood.

\section{The source of deviation}
\indent

 Either simulation circumstances are not appropriate or
something in analytical approach is not sufficiently reliable. In
the former, some possible questions are imaginable and we discuss
them to rule out this possibility.

One proposal is to do the simulation with more samples, because
the number of recurrent configurations is much more than something
like 2,000,000 $(\det \Delta \sim(3.21)^{L\times M})$ and so, our
samples cover only a small part of configuration space of system.
To rule out such a possibility we repeated simulation for a
variety of samplings from 100,000 to 10,000,000 samples and no
meaningful change was observed in the amount of deviation. In
fact, the ergodicity of model makes our work reliable. Furthermore
for each sampling, the initial recurrent configuration is
generated randomly but the behavior observed in all of them is the
same. Therefore this deviation cannot specific to some particular
parts of configuration space.

Other critique of our claim may be the difference between the
definition of probability of cluster in analytical approach and in
our simulations. In analytical method, probability is defined in
terms of the number of configurations in which the specific
cluster appears in a given place, but in our case it is defined in
terms of the number of appearance of the specific cluster in a
given configuration. Although this difference really exists, the
average of probability over all places in the former should be
equal to the average of probability over all recurrent
configurations in the latter. So this can not address our question
about the source of deviation. While this argument seems to be
enough, we measured probability of clusters $S_{8}$ and $S_{9}$
directly with the first definition in different areas. In all
cases we found the same difference as before between $P(S_{8})$
and $P(S_{9})$.

The other possible source of error could be finite size of the
lattice in simulations, as the analytical values have been
calculated exactly for infinite lattice. It is clear that for very
small lattices, the calculated quantities do differ noticeably
with results of infinite lattice. In larger latices, finite size
effect may also cause a little deviation from infinite case, but
this can not be the source of observed deviation. It seems that
the finite size effect is much smaller than the observed deviation
in our simulation, as it has not appeared in majority of clusters
in our simulation circumstances.

Another point about this critique is the fact that the similarity
of $S_8$ and $S_9$ does not concern only infinite lattice.
According to the analytical approach, the probability of
appearance of clusters $S_8$ and $S_9$ are equal in every lattice
and this claim of analytical approach is in considerable
disagreement with our simulation results.

Using simulation for lattices with different sizes from $5\times5$
to $2000\times2000$ shows that the difference between $P(S_{8})$
and $P(S_{9})$ not only does not decrease with size, but also
increases; whereas for lattice of size $5\times5$, the difference
between the two values is not statistically significant. But the
problem arises quickly and becomes more serious for larger
lattices specially because the error bar narrows down.

Finally, we may now focus on the analytical results. There does
not seem to be a calculational problem. Therefore we suspect that
a fundamental problem is involved which is however obscure to us.
\\

\begin{Large}
\textbf{Acknowledgments}
\end{Large}
\\

 We are grateful to S. Moghimi-Araghi and M. A.
Rajabpour for fruitful discussions.

\end{document}